\documentclass[prd,nofootinbib,
amsmath,amssymb,aps,twocolumn,
]{revtex4-1}

\usepackage[colorlinks=true
,urlcolor=blue
,anchorcolor=blue
,citecolor=blue
,filecolor=blue
,linkcolor=blue
,menucolor=blue
,linktocpage=true
,pdfproducer=medialab
,pdfa=true
]{hyperref}

\usepackage{hyperref}
\usepackage[T1]{fontenc}
\usepackage{graphicx}
\usepackage{comment}
\usepackage{bm}
\usepackage{braket}
\usepackage{cancel}
\usepackage{lmodern}
\usepackage{ulem}

\newcommand{\eq}[1]{\begin{equation}\begin{split} #1 \end{split}\end{equation}}

\newcommand{\lr}[1]{\left( #1 \right)}

\newcommand{\m}{\mathrm}

\begin{document}
\title{Population III star explosions and Planck 2018 data}
\author{Katsuya T. Abe}
\email{abe.katsuya@e.mbox.nagoya-u.ac.jp}
\author{Hiroyuki Tashiro}
\affiliation{Division of Particle and Astrophysical Science, Graduate School of Science, Nagoya University, Chikusa, Nagoya 464-8602, Japan}

\begin{abstract}
We investigate the effect of the population III (Pop III) stars supernova explosion~(SN) on the high redshifts reionization history using the latest Planck data.
It is predicted that massive Pop~III stars~($130M_\odot\leq M\leq 270M_\odot$) explode energetically at the end of their stellar life as pair-instability supernovae (PISNe).
In the explosion, supernova remnants grow as hot ionized bubbles and enhance the ionization fraction in the early stage of the reionization history.
This enhancement affects the optical depth of the cosmic microwave background~(CMB)
and generates the additional anisotropy of the CMB polarization on large scales.
Therefore, analyzing the Planck polarization data allows us to examine the Pop III star SNe and the abundance of their progenitors, massive Pop III stars.
In order to model the SN contribution to reionization, we introduce a new parameter $\zeta$, which relates to the abundance of the SNe to the collapse fraction of the Universe.
Using the Markov chain Monte Carlo method with the latest Planck polarization data, we obtain the constraint on our model parameter, $\zeta$.
Our constraint tells us that observed CMB polarization is consistent with the abundance of PISNe predicted from the star formation rate and initial mass function of Pop III stars in recent cosmological simulations.
We also suggest that combining further observations on the late reionization history such as high redshift quasi-stellar object~(QSO) observations can provide tighter constraints and important information on the nature of Pop III stars.
\end{abstract}
\maketitle

\section{Introduction}\label{Introduction}
From recent observations and theoretical studies, it is believed that the first stars known as population~III~(Pop~III) stars played essential roles in the history of the cosmological structure formation.
As the first luminous objects in the Universe, they formed around a few hundred million years after the big bang~(the redshift $z \sim 10$--$30$)~\cite{2002ApJ...564...23B}.
After their birth, Pop III stars contributed to the ionizing and heating of the surrounding intergalactic medium~(IGM) gas~\cite{2006ApJ...639..621A,2007ApJ...665...85J} and provided a significant impact on the first galaxy formation ~\cite{2012ApJ...745...50W,2013RvMP...85..809K}.
They could also trigger the formation of supermassive black holes~\cite{Venemans_2013,Wu_2015,2014AJ....148...14B}.
However, despite their importance, the detailed nature of Pop III stars is still unknown.
Various observational approaches are demanded to obtain further information about Pop III stars.

Although compared with typical stars at present, Pop III stars are luminous and massive, $m \gtrsim 10 ~M_\odot$~\cite{Hirano_2015,Susa_Hasegawa_2014}, it is difficult to observe them directly.
However, the recent studies pointed out that Pop III stars with a mass between $130M_{\odot}$ and $270M_{\odot}$ end with pair-instability supernovae~(PISNe), which is 
roughly $100$ times more powerful than typical Type Ia or Type II SNe~\cite{Herger_Woosley_star_nucleosynthetic,Umeda_2002}.
Furthermore, cosmological simulation~\cite{Hirano_2015,Susa_Hasegawa_2014} also show such relatively massive Pop III stars, and therefore their PISNe would not be rare.
Hence, it could be possible that we obtain the probe for the PISNe from the cosmological and astrophysical measurements.

One way to get the probe is the next-generation observation of near infrared. 
The redshifted ultraviolet emission from PISNe in high redshifts is a good target for it, such as the James Webb Space Telescope\footnote{https://www.jwst.nasa.gov/} and the Nancy Grace Roman Space Telescope\footnote{https://roman.gsfc.nasa.gov/}.
So far, there are a lot of theoretical works which examine the detectability of such PISNe using these observations~(e.g.~Refs.\cite{2005ApJ...633.1031S,2011ApJ...734..102K,Whalen_2012}).
Besides near-infrared observations,
it is also suggested that the sampling of metal-poor stars in the Milky Way
can provide the limit on the PISNe rate~\cite{2017MNRAS.465..926D}.

Additionally, Ref.~\cite{Peng_Oh_2003} studied the effect of 
PISNe in high redshifts on the temperature anisotropy 
of the cosmic microwave background~(CMB).
Since the gas inside an SN remnant~(SNR) is a hot ionized plasma,
CMB photons passing through the SNR suffer the inverse-Compton scattering.
That is the thermal Sunyaev-Zel'dovich~(tSZ) effect of PISNe, creating the CMB temperature anisotropy on small scales.
Although the anisotropy amplitude depends on the model of Pop III stars and PISNe,
they showed that the tSZ temperature anisotropy due to PISNe could be subdominant to the one from galaxy clusters.

This work investigates the effect for the global ionization fraction of PISNe in high redshifts with Planck polarization data.
The gas inside the SNRs of PISNe is compressed and fully ionized.
If many PISNe occur, the CMB photons suffer more scattering, and the E-mode angular power spectrum of CMB traces it.

Using Markov chain Monte Carlo(MCMC) method with the Planck 2018 polarization data, we constrain the amount of PISNe events.
After that, we also show that the restraints would lead us to the further astrophysical information of Pop III stars.

The rest of this paper is organized as follows. 
In Sec.~II, we describe the time evolution of the SNR shock shell.
Accordingly,
we show the relevant time scale for this work.
In Sec. III, introducing the effect for global ionization fraction due to the PISNe,
we explain our reionization model considered here.
After that, we show the equation of computing the number density of the PISNe with the model parameter.
In Sec. IV, we explain the MCMC methods used in this work and show the resulting constraint. Subsequently, we discuss the restriction compared with the cosmological simulation about the Pop III stars in Sec.V. 
Finally, we summarize in Sec.~VI.
Throughout our paper,
we take the flat $\Lambda$CDM model with the Planck best fit parameters~\cite{Planck2018_cospara}:
$(\Omega_{\rm m},\Omega_{\rm b},h,n_{\rm{s}},\sigma_{8})$=$(0.32,0.049,0.67,0.97,0.81)$. 

\section{The properties of Supernova remnants of Pop III stars}
\label{section2}
Since Pop III stars are massive, $m \gtrsim 10 ~M_\odot$~\cite{Hirano_2015,Susa_Hasegawa_2014}, it is theoretically predicted that
Pop III stars cause SNe at the final stage of their lives, which is about 1$\rm{Myr}$ after its birth.
In addition, from the recent studies, the Pop III stars with mass between $130M_{\odot}$ and $270M_{\odot}$ end with super energetic SNe, called PISNe, which are 
roughly $100$ times more powerful than typical Type Ia or Type II SNe~\cite{Herger_Woosley_star_nucleosynthetic,Umeda_2002}.
Once supernovae occur, the supernovae remnants~(SNRs) would expand with a shock wave. In this section, we describe the time evolution of the general SNR with the analytical model.

After occurring the SN explosion,
a certain mass is ejected into a surrounding gas with supersonic velocity.
The ejecta sweeps up the surrounding gas, creating the expanding shock waves.
This is a trigger to form the SNR.
The SNR expands outwards nearly spherically.

The evolution of the SNR has mainly three phases~\cite{Reynolds_2017}. The first phase is called the free-expansion phase.
In this initial phase,
the swept-up mass by the SNR is negligible compared with the ejected mass. Therefore, the evolution of the SNR in this phase is determined by only the initial energy and the ejected mass.
The SNR evolution enters the second phase, the adiabatic phase, when
the mass of the swept-up surrounding gas is comparable with
the initial ejected mass.
The swept-up surrounding gas 
is compressed and heated by the shock
and
forms a shell structure. 
The evolution in this phase
is well described by
the Sedov-Taylor self-similar solution.
As the SNR evolves, the velocity of the SNR decreases, and the resultant expansion times scale of the SNR becomes long. 
Finally,
since the expansion time scale will be longer than the cooling time scale, 
the radiative cooling is not negligible in the evolution of the SNR.
This third phase is called
the momentum conserving phase.
The thermal energy of the SNR is lost by the radiative cooling. 
The expansion of the SNR just followed the momentum conservation.

To evaluate the impact of the SNR as the cosmological ionization photon source,
we are interested in the second phase, the adiabatic phase.
This is because
the first phase has a very short duration and,
in the third phase, 
most energy is taken away
to the CMB through the inverse-Compton scattering.

As mentioned above, the evolution of the SNR shocked shell in the adiabatic phase is well described by the Sedov-Taylor self similar solution.
In this solution, the radius evolution of the shocked shell can be written as the function of the SN explosion energy $E_{\rm{SN}}$:
\begin{equation}\label{rsn}
R_{\rm{SN}}(t)=2.0~[\mathrm{kpc}] \left[
\left(\frac{t}{10^{7} \mathrm{yr}}\right)^{2 }
 \left(\frac{E_{\mathrm{sn}}}{10^{46} \mathrm{J}}\right)\left(\frac{10^{3} \mathrm{m}^{-3}}{n_{\rm{g}}}\right)
 \right]^{\frac{1}{5}},
\end{equation}
where $t$ represents the time after the SN explosion, and $n_{\rm{g}}$ is the
number density of the hydrogen atom in the outer gas of the shocked shell. 
First the SNR propagates in a denser gas in the host dark matter halo and subsequently in the IGM outward. 

In this work, we neglect the effect of the overdensity in a halo, and set to $n_{\m{g}}\approx n_{\m{b,IGM}}$, where $n_{\m{b,IGM}}$ is the number density of baryons in the IGM.
Although the SNR can expand larger than the virial radius,
high density gas in a halo reduce the energy of the SNR in the IGM propagation and decrease the radius given in Eq.~\eqref{rsn}.
In order to evaluate such an overdensity effect, one needs to perform the numerical calculation including the density profile in where the SNR propagates.

In the limit of a strong shock,
the number density in the shell,~$n_{\rm SN}$, is
related to the surrounding one~$n_{\mathrm{g}}$ with
the adiabatic index $\gamma$, $n_{\rm{SN}}=(\gamma+1)/(\gamma-1)n_{\mathrm{g}}$.
Furthermore, the thickness of the shocked shell,~$\Delta R_{\rm SN}$, is
obtained from the mass conservation law as 
$\Delta R_{\rm{SN}}(t)=(\gamma-1)/(3(\gamma+1))R_{\rm{SN}}(t)$.
Here, we neglect the density profile in the shock shell. 
For $\gamma$, we adopt a monoatomic gas case, $\gamma=5/3$~\cite{Barkana_2001}.

The adiabatic phase terminates 
when the cooling becomes effective.
Since the gas in SNRs is fully ionized by the shock heating,
the major cooling mechanism is 
Compton cooling.
The time scale of Compton cooling is given by
\begin{equation}
t_{\mathrm{C}}=\frac{3m_{\m{e}}}{4\sigma_{\mathrm{T}}aT^4_{\mathrm{\gamma}}}=1.4\times 10^7\lr{\frac{1+z}{20}}^{-4}\mathrm{yr}.
\end{equation}
The SNR evolves following the equation~\eqref{rsn} until $t=t_{\rm C}$.
After that, the thermal energy, which drives the shell expansion, is quickly lost by Compton cooling.
In this paper, we simply 
evaluate the effect of SNRs discussed in the following section at $t=t_{\rm C}$.

The radial profiles of the electron density in an SNR is given by
\begin{equation}
n_{\rm e}(r) = 
    \frac{\gamma+1}{\gamma-1}n_g, \quad 
    \lr{\frac{\gamma-1}{3(\gamma+1)}r_{\m{SN}}<r<r_{\m{SN}}}
\end{equation}
where $r$ is the comoving radial distance from the center of the SNR and, therefore,  
$r_{\rm SN}$ is $r_{\rm SN} = (1+z)R_{\rm SN}$.
As the SNRs are cooled,
electrons in them are recombined again.
Accordingly, the effect for global ionization fraction from the SNR is suppressed.
The time scale of recombination in the SNR can 
be written as
\begin{equation}
t_{\rm rec} = \frac{\gamma-1}{\alpha_{\m{B}}(t_{\m{C}})(\gamma+1)n_{\m{g}}}=3.3\times 10^7\lr{\frac{1+z}{20}}^{-0.12}\m{yr},
\end{equation}
where $\alpha_{\rm{B}}$ is the case B recombination rate given in Ref.~\cite{Fukugita&Kawasaki_cooling}.
This time scale is not negligible, compared with the cosmological time scale,~$t_{\m{cos}}$. 
We take into account this suppression in the abundance of PISNe in the next section.

\section{THE REIONIZATION MODEL}
In the standard analyses of the reionization history adopted by Planck CMB measurements, only the overall optical depth of electrons is considered assuming a $\mathrm{tanh}$-model reionization history.
The polarization data, however, should contain additional information for the full reionization history.
Here we investigate the effect of Pop III star supernovae especially in PISNe for the global ionization history with these data.

\subsection{Reionization model}
In the reionization models considered here, we add the effects from PISNe of Pop III stars to the fiducial ionization history adopted by Planck CMB measurements.
We assume that the Pop III stars are only hosted by the massive halos
with the virial temperature $T_{\m{vir}}>10^4~\m{K}$.
The condition of $T_{\mathrm{vir}}>10^4~\mathrm{K}$ comes from the efficiency of the atomic cooling in the halo.

It is a fact that the Pop III stars can be formed in the halos which are not satisfied by this condition.
However, if the virial temperature $T_{\mathrm{vir}}$ is lower than $10^4\mathrm{K}$, the star formation rate is suppressed and even in the halo-host-star case, may become "one star per halo" because internal UV photodissociation of $H_2$ by the Pop III stars ceases further gas cooling and star formation~\cite{1999ApJ...518...64O}. 
Moreover, in the case of the more massive halos with $T_{\mathrm{vir}}>10^4\mathrm{K}$, there is a conceivable scenario that many stars form together in such a halo where atomic cooling allows gas to collapse and have much higher density~\cite{2002ApJ...569..558O}.

The effect of PISNe on the cosmic reionization could be subdominant
and the main reionization photon sources are Pop II stars and first galaxies. 
Therefore, taking into account the PISNe reionization effect,
we assume that the evolution of the global ionization fraction can be decomposed into three terms, 
\eq{\label{eq: ion_model}
x_e(z)=x_e^{\m{rec}}(z)+x_e^{\m{reio}(z)}+x_e^{\m{SN}}(z),
}
where $x_e^{\m{rec}}$ is the global ionization fraction in the recombination epoch and $x_e^{\m{reio}}$ represents 
the contribution from the main reionization source including Pop II stars and galaxies.
For obtaining $x_e^{\m{rec}}$, we employ the recombination code~{\tt RECFAST}~\cite{1999ApJ...523L...1S,2000ApJS..128..407S,2008MNRAS.386.1023W,2009MNRAS.397..445S}.
Then, we adopt the widely used "tanh" model for $x_e^{\m{reio}}$~\cite{2008PhRvD..78b3002L},
\begin{eqnarray}
 x_e^{\m{reio}}(z)&=&x_e^{\mathrm{before}}+\frac{1}{2}\left(x_e^{\m{after}}-x_e^{\mathrm{before}}\right)
 \nonumber \\
&&\quad \
 \times \left[1+\tanh \left(\frac{y^{\mathrm{reio}}-y(z)}{\Delta y}\right)\right],
 \label{eq:tanh-shape}
\\
y(z)&=&(1+z)^{3 / 2},
\end{eqnarray}
where
$y^{\m{reio}}=y(z^{\m{reio}})$, $\Delta y=1.5 \sqrt{1+z^{\m{reio}}} \Delta z$ with the duration of reionization, $\Delta z=0.5$. 
In Eq.~\eqref{eq:tanh-shape},
$x_{\mathrm{e}}^{\mathrm{after}}$ is the ionization fraction after finishing reionization, $x_{\mathrm{e}}^{\mathrm{after}}$=1 and $x_{\mathrm{e}}^{\mathrm{before}}$ is the left-over ionization fraction well after the recombination epoch adopted as $x_{\mathrm{e}}^{\mathrm{before}}=10^{-4}$.

The impact of PISNe on the reionization process is provided by the additional term, $x_e^{\m{SN}}$.
Since the gas inside SNRs is fully ionized, the volume occupation of the SNRs represents the global ionization fraction.
Thus we estimate the SN term by
\eq{\label{eq: ion_add}
x_e^{\m{SN}}(z)= f_{\m{ion}}(z)n_{\m{SN}}(z)V_{\m{ion}}(z), 
}
where $f_{\rm ion}(z)$, $n_{\m{SN}}$ and $V_{\m{ion}}$ represent the survival probability of
ionized SNRs, the number density of PISNe, and the volume of each ionized SNR respectively.
In this form of additional ionization fraction of Eq.~\eqref{eq: ion_add}, we assume that each SNRs cover a different region.
Although it is totally ionized soon after the creation,
the inside of SNRs gradually become
neutral in the time scale of recombination,
$t_{\m{rec}}$.
In order to account for this effect, we introduce the probability 
$f_{\m{ion}}(z)=t_{\m{rec}}(z)/t_{\m{cos}}(z)$
with the upper bound, $f_{\m{ion}}\leq 1$.
%For $f_{\rm ion}(z)$, to take into account the recombination timescale $t_{\m{rec}}$, we take $f_{\m{ion}}(z)=t_{\m{rec}}(z)/t_{\m{cos}}(z)$
%with the upper limit $f_{\m{ion}}\leq 1$.
The volume $V_{\m{ion}}$ is given by $V_{\m{ion}}(z)=4\pi/3 R_{\m{SN}}^3(t_{\m{c}},z)$ using the radius the SNe in ~\eqref{rsn} with $E_{\m{sn}}=10^{46}\m{J}$.
In the next subsection, we discuss the number density of PISNe, $n_{\rm SN}$.

In our model, we assume that each PISN occurs isolatedly and an ionized SNR expands in the neutral IGM to increase the ionization fraction.
This assumption could lead to overestimate the contribution of SNRs to the ionization fraction.
In Sec.~\ref{subsec: limit_assumption}
we will discuss the limitation of our assumptions and 
the cases where our assumption is not applicable.

\subsection{The abundance of PISNe }
Since the abundance of PISNe has not been well decided yet because of lots of theoretical uncertainties (i.e. the mass function of the Pop III stars), here we consider it is proportional to the collapsed mass of baryon in dark matter halo.
We model the number density of PISNe at given $z$ as
\eq{\label{eq: numsn}
n_{\rm SN}(z) = \zeta \frac{1}{m_*} f_{\m{coll}}(M_{\m{min}})\bar{\rho}_{\m{b}}(z),
}
where $\zeta$ is the model parameter
whose combination $\zeta f_{\m{coll}}\bar{\rho}_{b}$ means the total mass of the Pop III stars which occurs PISNe in one halo, $M_{\m{min}}$ is the mass corresponding to $T_{\m{vir}}$, $\bar{\rho}_{\m{b}}$ is the background baryon density, and $m_*$ is the typical mass of the Pop III star which occurs PISNe. Although it is known that the Pop III stars cause PISNe in the case of mass range $[130\m{M}_{\odot},270\m{M}_{\odot}]$\cite{Herger_Woosley_star_nucleosynthetic}, we simply assume $m_*=130\m{M_{\odot}}$ in our model.
We set the geometry of the gravitational collapse to spherical one (i.e. the halo mass function is Press-Schechter). Then, $f_{\m{coll}}(M)$ which is the collapse fraction in halos with the mass $M_{\m{halo}}>M$ is calculated by
\eq{
f_{\mathrm{coll}}(M)=\frac{2}{\sqrt{2 \pi} \sigma(M)} \int_{\delta_{c}}^{\infty} d \delta 
\exp  \left(-\frac{\delta^{2}}{2 \sigma^{2}(M)}\right)
%\exp  \left(-\frac{\nu^{2}}{2 }\right)
= \operatorname{erfc}\left(\frac{\nu}{\sqrt{2}}\right),
}
where $\nu\equiv\delta_c/\sigma(M)$ and $\delta_c=1.67$.
The variance of the matter density fluctuation~$\sigma$ is written by
\eq{\label{eq: dens_variance}
\sigma^2(M) = \int \m{dlog}k~ W^2(kR_{\m{vir}})\mathcal{P}(k),
}
where $R_{\m{vir}}$ is the virial radius for $M$. Here $W(kR)$ is the 3D window function. In this work, we employ the top-hat window function
\eq{
W\left(k, R\right)=\frac{3}{\left(k R\right)^{3}}\left(\sin \left(k R\right)-kR\cos \left(k R\right)\right).
}
The nondimensional matter power spectrum $\mathcal{P}(k)$ can be calculated as
\eq{
\mathcal{P}(k)=\frac{4}{25}\lr{\frac{(1+z)k}{H_0}}^4T_{\m{q}}^2~\mathcal{P}_{\mathcal{R}}(k),
}
using the transfer function $T_{\m{q}}$ formulated by Bardeen et al.~\cite{1986ApJ...304...15B},
\eq{
T_{q}=&\frac{\ln (1+2.34 q)}{2.34 q}\\
&\ \times\left[1+3.89 q+(16.1 q)^{2}+(5.46 q)^{3}+(6.71 q)^{4}\right]^{-1 / 4},
}
where $q \equiv k/\Gamma~h \m{Mpc}^{-1}$, and $\Gamma$ is the apparent shape parameter including baryonic effect~\cite{1995ApJS..100..281S}, $\Gamma\equiv \Omega_{\m{m}}h\m{exp}(-\Omega_{\m{b}}-\sqrt{2h}\Omega_{\m{b}}/\Omega_{\m{m}})$.
The nondimensional primordial power spectrum $\mathcal{P}_{\mathcal{R}}$ is 
\eq{
\mathcal{P}_{\mathcal{R}}=\mathcal{A}_s\lr{\frac{k}{k_{\m{pivot}}}}^{n_s-1},
}
where $\mathcal{A}_s$, $k_{\m{pivot}}$ and $n_s$ are the amplitude of the primordial scalar power spectrum, the pivot scale, and the scalar spectral index respectively.

As Pop III star formation proceeds, the primordial IGM is
contaminated by metals through SNe of Pop III stars.
When the metallicity reaches the critical threshold value
at $z_{\rm end}$,
the formation of Pop III stars terminates.
Although new Pop III PISNe no longer happen after that,
SNRs created until $z_{\rm end}$ still survive for a while because the recombination time scale in SNRs is
comparable the cosmological time scale at that redshift.
Therefore, SNRs of Pop III stars can contribute the 
global ionization fraction even after $z_{\rm end}$
In order to take this contribution, we provide $x_e^{\m{SN}}(z)$ as 
\eq{
x_e^{\m{SN}}(z)= f_{\rm ion }(z)n_{\m{SN}}(z_{\m{end}})V_{\m{ion}}(z_{\m{end}})~\quad (z<z_{\m{end}}),
}
where we assume that the SNRs created at $z_{\rm end}$
fade away in the time scale $t_{\m{rec}}(z)$.
For simplicity, we set $z_{\m{end}}=12$ in this work.
We will discuss the impact of $z_{\rm end}$ on our analysis later.

\begin{figure}
\centering
\includegraphics[width=8cm,clip]{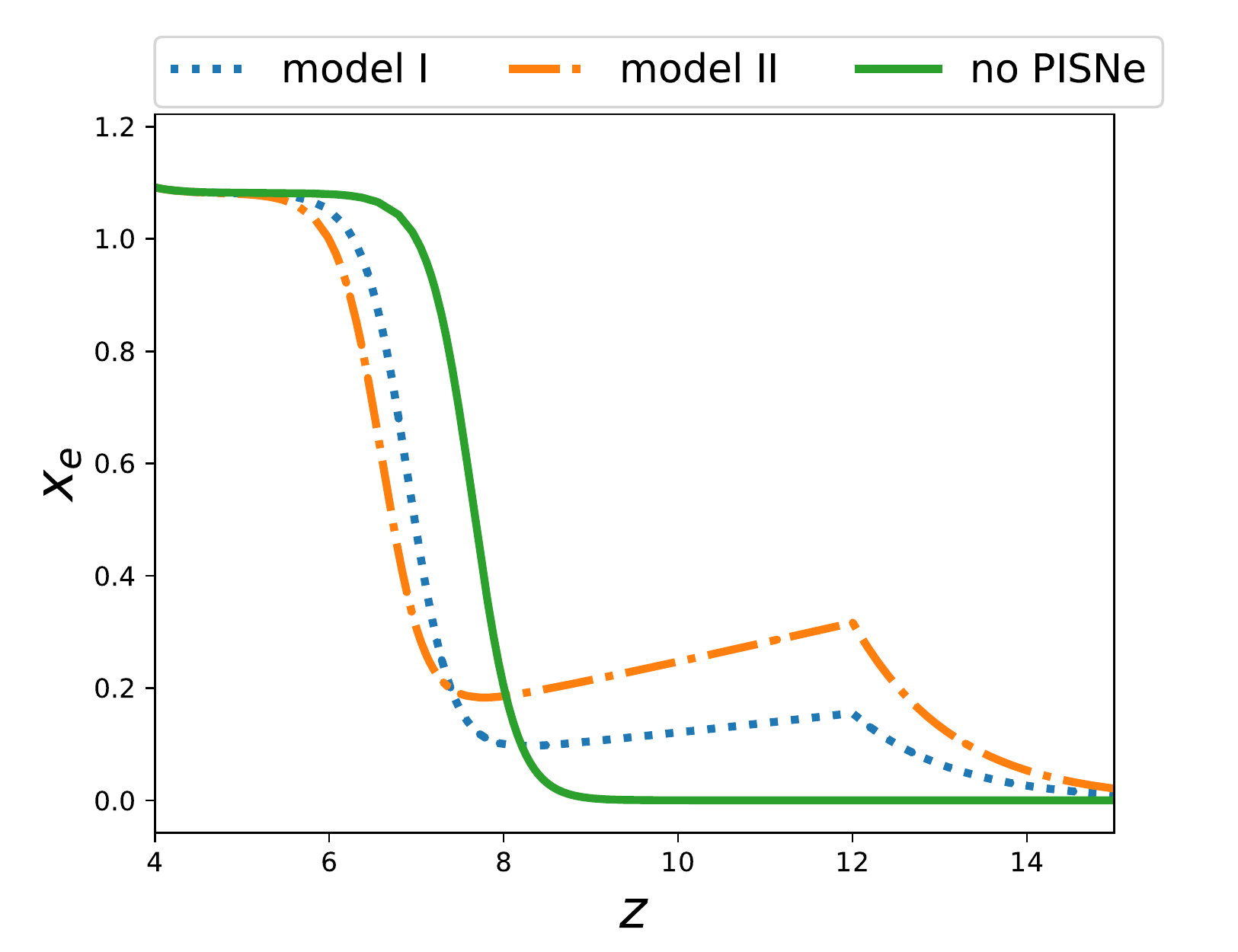}
\caption{global ionization history with several values of $\zeta$.}
\label{fig: xe_zeta}
\end{figure}

Figure~\ref{fig: xe_zeta} shows the global ionization history with Pop III PISNe models.
In the model~I and II, we set the model parameter to
$(z_{\m{reio}},\m{log}_{10}\zeta)=(6.90,-2.17),(6.60,-1.86)$, respectively. 
For comparison, we plot the standard reionization model without the PISNe effects.
One of good indicators for the cosmological reionization history is the optical depth of the Thomson scattering for CMB photns, 
\eq{\label{eq: optical_depth}
\tau = \int \frac{d z}{H(z)} \sigma_{\rm T } x_e(z) n_e(z) .
}
All of the three models
have the same optical depth of the Thomson scattering,
$\tau\simeq0.054$,
which is consistent with the Planck result.
We can see that the Pop III PISNe can enhance the ionization fraction in the early universe, $z_{\m{reio}}<z\leq 15$.

\if0
In our model, we neglect the reionization contribution from the progenitor Pop III stars of PISNe. During the stellar stage before PISN explosions, they can emit ionizing photons to the IGM and make ionized bubbles. However, the size of ionized bubbles depends on the ionizing photons' escape fraction,~$f_{\rm{esc}}$. Although there is still a large theoretical uncertainty in the escape fraction, some theoretical works predict the escape fraction smaller than the unity. For example, Ref.~\cite{2000ApJ...545...86W} reported that $f_{\rm{esc}}\lesssim \mathcal{O}(10^{-2})$ is preferred in the high redshifts from their simulations and Ref.~\cite{2020MNRAS.498.2001M} suggest the value of $f_{\rm{esc}}$ would be  $0.05<f_{\rm{esc}}<0.3$ at a high redshift $z\sim 10$.
From a simple estimation,
if the escape fraction of the progenitors of PISNe is $f_{\rm{esc}}\lesssim 0.3$,
we found that the Str\"{o}mgren sphere of the ionized bubbles are smaller than 
the SNR given in Eq.~\eqref{rsn}.
Therefore, in this paper, we assume that
the bubble created by a Pop III star before the SNe
can be destroyed by an SNR and negligible as a reionization source.

However, in general,  some fraction of Pop III stars can be progenitors of PISNe and the rest of them is not.
Therefore,
the effect of Pop III stars cannot be negligible, depending on the initial mass function of Pop III stars which is still under debate.
In this paper, although we ignore the contribution of Pop III stars on the cosmic reionization for simplicity,
we will come back to this issue in Sec~\ref {sec:results}.
\fi

In our model, we neglect the reionization due to the Pop III stars, which do not have enough mass to occur PISNe, although they also contribute to the early stage of the cosmic reionization. 
The fraction of such low-mass stars depends on the initial mass function of Pop III stars which is still under debate. 
In this paper, we ignore the contribution of Pop III stars on the cosmic reionization for simplicity. however, we will come back to this issue in Sec.~\ref{subsec: limit_assumption}.

%However, in general,  some fraction of Pop III stars can be enough massive to become progenitors of PISNe while the rest of them is not.
%Therefore 
%the effect of Pop III stars cannot be negligible, depending on the initial mass function of Pop III stars which is still under debate.

\section{MCMC analysis with Planck 2018}

In order to constrain the effect of PISNe based on our model in Eq.~\eqref{eq: ion_model}, we employ the MCMC analysis with Planck 2018 data.
Chains of MCMC samples are generated by the publicly open code {\tt MontePython}~\cite{Audren:2012wb}, which adopts the code {\tt CLASS}~\cite{Blas_2011} for calculating the theoretical CMB angular power spectrum.
We have modified the {\tt CLASS} code including the PISNe effect for global ionization fraction represented in Eq.~\eqref{eq: ion_model}.

The optical depth is mainly constrained by the reionization bump that appeared on small scales in the CMB polarization. 
Since we are interested in 
$\zeta$ and $z_{\m{reio}}$,
which mainly control the ionization history 
and the optical depth $\tau$ with $z^{\rm reio}$ in equation~\eqref{eq:tanh-shape},
we fix other cosmological parameters to the Planck best-fit parameter of the TT, TE, EE, low-$\ell$ + lensing
measurement,
$\Omega_{\m{b}}=2.237$, $\Omega_{\m{cdm}}=0.1200$, $100\theta_{\m{s}}=1.04092$, $\m{ln}10^{10}A_{\m{s}}=3.044$, and $n_s=0.9649$.
These parameters do not affect the reionization bump much.

To obtain accurate results from MCMC methods, it is essential to check if the MCMC chains contain enough samples which are independent of each other and cover a sufficient volume of parameter space such that the density of the samples converges to the actual posterior probability distribution. 
Therefore, here,
we run the MCMC chain until the Gelman and Rubin convergence statistic R, which represents the ratio of the variance of parameters between chains to the variance within each chain, satisfies $R-1<0.05$ ~\cite{1992StaSc...7..457G,doi:10.1080/10618600.1998.10474787}. 

\section{Results and Discussion}\label{sec:results}
 
\begin{figure}
\centering
\includegraphics[width=8cm,clip]{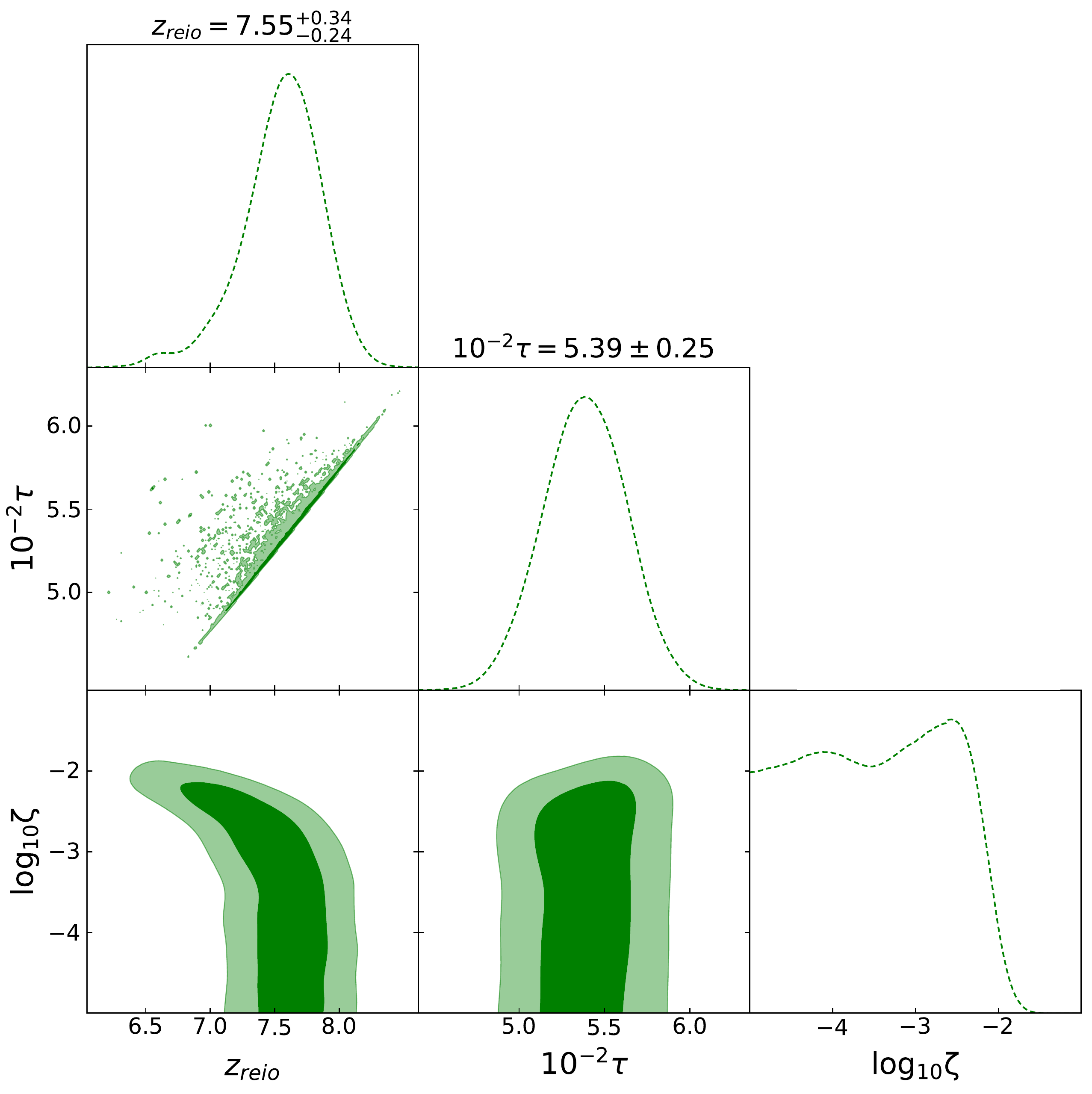}
\caption{MCMC result}
\label{fig: mcmc}
\end{figure}

Our resulting constraint is shown in Fig.~\ref{fig: mcmc}, in which 
$\zeta$ and $z_{\rm reio}$ are our model free parameters and the optical depth $\tau$ is derived from Eq.~\eqref{eq: optical_depth} with the sampling data of 
$\zeta$ and $z_{\rm reio}$.
The dark green region shows the $1\sigma$ region and the light green region represents the $2\sigma$ region.
Since the CMB anisotropy is sensitive to the total optical depth $\tau$ during and after the cosmic reionization, the Planck measurement basically provides the constraint on $\tau$. 
In our model, the main contribution to $\tau$ comes from the "\rm{tanh}" term while, the PISNe effect is subdominant.
Therefore, $z_{\m{reio}}$ for the "\rm{tanh}" term is strongly constrained.
When $\zeta$ increases more than $\zeta > 10^{-3}$, SNRs can induce early reionization and make a non-negligible contribution to $\tau$.
To compensate for this effect, small $z_{\m{reio}}$ is preferred as $\zeta$ becomes large as shown in Fig.~\ref{fig: mcmc}.
However, when
$\zeta$ is larger than $10^{-2}$, even only PISNe
can fully ionize the Universe.
Therefore, $\zeta > 10^{-2}$ can be ruled out.

The Planck measurement gives the constraint 
on our model parameter, $\zeta\leq10^{-2}$. 
Now let us discuss what implication on the physics of Pop III stars we can obtain from our constraint.
Our model parameter $\zeta$ is introduced to
connect between the number density of the PISNe and the collapse fraction as shown in  
Eq.~\eqref{eq: numsn}.
On the other hand, 
conventionally, one can relate the 
PISN density to the dark matter mass function 
\begin{align}\label{eq:number_density_SN}
&n_{\rm SN}(z) = \int_{m_{\m{min}}} ^{m_{\m{max}}} dm ~\frac{dn_{\ast} (m,z) }{dm},
\\
&\frac{dn_{\ast}}{dm}= \frac{g_\ast(m)}{m} \int_{M_{\m{min}}}\hspace{-3mm} dM ~ f_{\rm host}(M) f_{\rm \ast}(M) M\frac{dn (M,z)}{dM},
\label{eq:starfunc}
\end{align}
where $dn_{\ast} (m,z) /dm$ is the mass function of Pop III stars with a mass $m$ at a redshfit $z$, and $m_{\m{min}}$ and $m_{\m{max}}$ are the lower and upper mass limit of the Pop III stars which occur PISNe.
In Eq.~\eqref{eq:starfunc}, 
$dn(M,z)/M$ is the mass function of dark matter halos with a halo mass $M$ at $z$,
$M_{\rm min}$ is a minimum mass of dark matter halos for hosting Pop III stars,
$f_{\rm host}(M)$ is
the fraction of dark matter halos with mass $M$
which can host stars, $f_{\rm \ast}$ is the fraction of the total stellar mass to the dark matter halo mass~$M$, and $g_\ast(m)$ is the initial mass function~(IMF) of Pop III stars which is normalized as $\int dm~g_{\ast}(m) =1$~($g_\ast$ has a dimension of $(\rm mass)^{-1}$). 
In general, $f_{\mathrm{host}}$, $f_{\rm \ast}$ and $g_\ast (m)$ also depend on redshift $z$.

For the mass function of dark matter halos, $dn(M)/dM$, we adopt the Press-Schechter theory here.
Therefore, we can relate the mass function with collapse fraction $f_{\m{coll}}$ as
\eq{
f_{\m{coll}}\bar{\rho}_{\m{m}}(z) =\int_{M_{\m{min}}}dM~M\frac{dn_{\ast} (m,z) }{dm},
\label{eq:fcoll_def}
}
where $\bar{\rho}_{\m{m}}(z)$ is the background matter density at $z$.
It is useful to define the weighted average value for 
$f_{\rm host}(M)$ and $f_{\rm \ast}(M)$
\eq{\label{def: mathcalB}
\bar{f}_{\m{X}}=\frac{\int_{M_{\m{min}}}\hspace{-3mm} dM ~ f_{\m{X}}(M) M\frac{dn (M,z)}{dM}}{\int_{M_{\m{min}}}\hspace{-3mm} dM ~M\frac{dn (M,z)}{dM}},
}
where the subscript~$\m X$ stands for ${\ast}$ or $\m{host}$.
We also introduce the number fraction of PISN progenitors to total Pop III stars as
\begin{equation}\label{def: fmf}
f_{\m{mf}} \equiv \int_{m_{\rm min}} ^{m_{\rm max}} dm ~g_\ast(m).
\end{equation}
If the IMF is the delta-function type mass function, $g_\ast(m)=\delta_{\m{D}}(m-m_*)$ with ${m_{\rm min}}< m_* <{m_{\rm max}}$, $f_{\m{mf}}$ equals to unity, and if the IMF is the mass function obtained from Pop III star formation simulation in Ref.~\cite{Susa_Hasegawa_2014}, it is about $g_\ast(m)\sim 0.3$.
Here we set $(m_{\m{min}}, m_{\m{max}}) = (130\m{M}_{\odot},270\m{M}_{\odot})$ as before.

Using Eqs~\eqref{eq:fcoll_def}~\eqref{def: mathcalB} and \eqref{def: fmf}, we can approximately estimate
the number density of PISNe from Eq.~\eqref{eq:number_density_SN} in
\eq{\label{eq: num_sn_john}
n_{\rm SN}(z) 
\approx \frac{1}{m_*}f_{\m{mf}}
f_{\rm star}
f_{\m{coll}}(M_{\m{min}})\bar \rho_{\m{m}}(z),
}
where $f_{\rm star}$ is defined as $f_{\rm star} \equiv \bar{f}_{\m{host}}\bar{f}_{*}$ and
represents the fraction of the total stellar mass to the total dark matter halo mass in the universe.
Comparing both Eqs.~\eqref{eq: numsn} and \eqref{eq: num_sn_john}, we obtain the relation as
\eq{\label{relation:zeta}
\zeta\approx f_{\m{mf}}f_{\rm star}\frac{\Omega_{\m{m}}}{\Omega_{\m{b}}}.
}
Therefore, the constraint, $\zeta \lesssim 10^{-2}$, can be converted into 
\eq{\label{eq: main_result}
f_{\m{mf}}f_{\rm star} \lesssim 1.4\times10^{-3}.
}

Cosmological numerical simulations suggest {$f_{\rm star}\lesssim 10^{-3}$} around the epoch of Pop III star formation in Ref.~\cite{2014MNRAS.442.2560W},
although there are still some uncertainties in both our theoretical model and the redshift evolution of $f_{\rm star}$~($\bar{f}_{\m{host}}$ and $\bar{f}_{*}$).
Therefore, it is difficult to provide the constraint on $f_{\m{mf}}$ 
from our MCMC analysis on $\zeta \lesssim 10^{-2}$.
However, it is worth mentioning that, if further observations provide more information on the evolution of the ionization fraction during reionization, the constraint on PISNe allows us to access the Pop III star IMF through $f_{\m{mf}}$. For example, the recent high-redshift quasi-stellar object~(QSO) observation suggests that the volume-averaged neutral fraction is 
$\langle x_{\rm HI} \rangle ={0.60} $ at $z = 7.54$. When considering this result, our constraint could be improved to $\zeta \lesssim 10^{-3}$~\cite{2018ApJ...864..142D}.   
In this case, our constraint tells us $f_{\rm mf} < 0.1$ and 
prefers the Pop III star IMFs in which the progenitors of Pop III stars are subdominant in the terms of the total Pop III star abundance.

In our model, one of the most important uncertainties is 
$z_{\m{end}}$, which is the redshift for the termination of PISNe.
In general, $z_{\m{end}}$ is significantly related to the metal pollution of the Universe, that is, the cumulative number density of PISNe. However, in this paper, we introduce $z_{\m{end}}$ by hand. In order to investigate the impact of $z_{\m{end}}$ on the constraint of $\zeta$, we perform the MCMC analysis with different $z_{\m{end}}$ between $10 < z_{\m{end}}<14$.
As a result, our constraint is changed by about $\pm25\%$ and
we find out the fitting form in
\eq{
\m{log}_{10}\zeta \leq -2.0\lr{\frac{z_{\m{end}}}{12}}^{1.22}.
}
The second one is the energy injected into SNRs of PISNe, $E_{\m{sn}}$. 
In this paper, although we adopt a constant injected energy, $E_{\m{sn}}=10^{46}\m{J}$,
it depends on the progenitor mass and the metallicity.
In our model, $E_{\m{sn}}$ affect our constraint through the SNR volume in Eq.~\eqref{eq: ion_add} where 
one can see that both $\zeta$ and $E_{\m{sn}}$ degenerate each other.
Therefore, our constraint on $\zeta$ have the dependence on $E_{\m{sn}}$,
\eq{
\zeta \leq 10^{-2}\lr{\frac{E_{\m{sn}}}{10^{46}\m{J}}}^{-3/5} .
}

In this paper, we neglect the effect on the reionization process, which Pop III stars provide directly by emitting the ionization photons during their main sequence.
%\KA{assuming their $f_{\rm{esc}}\lesssim 0.3$ as Refs.~\cite{2000ApJ...545...86W,2020MNRAS.498.2001M} suggest}. 
The authors of Ref.~\cite{Miranda2017} have investigated this effect on the early stage in the reionization history.
They parametrized the abundance of Pop III stars, relating the collapse fraction as we have done for the parametrization of the PISN abundance in this paper
, and provide the constraint by using MCMC methods with Planck 2015 data.
Using the similar way to obtain Eq.~\eqref{eq: main_result},
their result suggests  $\bar f_{\m{esc}} f_{\rm star}\leq 10^{-2}$ where $\bar f_{\m{esc}}$ is the weighted average escape fraction of ionizing photon for dark matter halos.
Therefore, the constraints on PISNe and Pop III stars are complementary:
the constraint on PISNe is sensitive to the IMF of Pop III stars through $f_{\rm mf}$ while the one on Pop III stars provides useful information on $\bar f_{\m{esc}}$.

\subsection{the limitation of our isolated SNR assumption}\label{subsec: limit_assumption}

In our model, we take the assumption that isolated PISNe create the SNRs expanding in the neutral IGM and increase the ionization fraction.
For the validity of this assumption, there are mainly two concerns. One is the ionized bubble created by a massive Pop III star before a PISN and the second is the overlapping (or clustering) of SNRs.
Before PISN explosions, massive Pop III star emit ionizing photon and creates the ionized bubbles.
When an ionized bubble is larger than a SNR of PISN, PISNe
cannot increase the ionization fraction and most of PISNe energy is consumed to heat up the SNRs.
The size of the ionized bubbles is roughly estimated by the Str\"{o}mgren radius,
$r_{\m{s}}$ which is given by the equilibrium between the number of ionizing photons and the neutral hydrogen.
With the ionizing photons emitting from a massive Pop III star, $N_\gamma$,
the Str\"{o}mgren radius in the IGM density is 
\eq{\label{eq: rion_with_z}
r_{\rm{s}} = 2.8\left[\left(\frac{f_{\rm{esc}}}{0.1}\right) \left(\frac{N_{\gamma}}{10^5}\right)\right]^{1/3}\hspace{-1mm}\left( \frac{13}{1+z}
\right)~\rm{kpc}.
}
where $f_{\rm esc}$ is the escape fraction of ionizing photons.
Although there is still a large theoretical uncertainty in the escape fraction, $f_{\rm esc}$, some theoretical works predict the escape fraction smaller than the unity. For example, Ref.~\cite{2000ApJ...545...86W} reported that $f_{\rm{esc}}\lesssim \mathcal{O}(10^{-2})$ is preferred in the high redshifts from their simulations and Ref.~\cite{2020MNRAS.498.2001M} suggest $0.05<f_{\rm{esc}}<0.3$ in a redshift $z\sim 10$.

\begin{figure}
\centering
\includegraphics[width=8cm,clip]{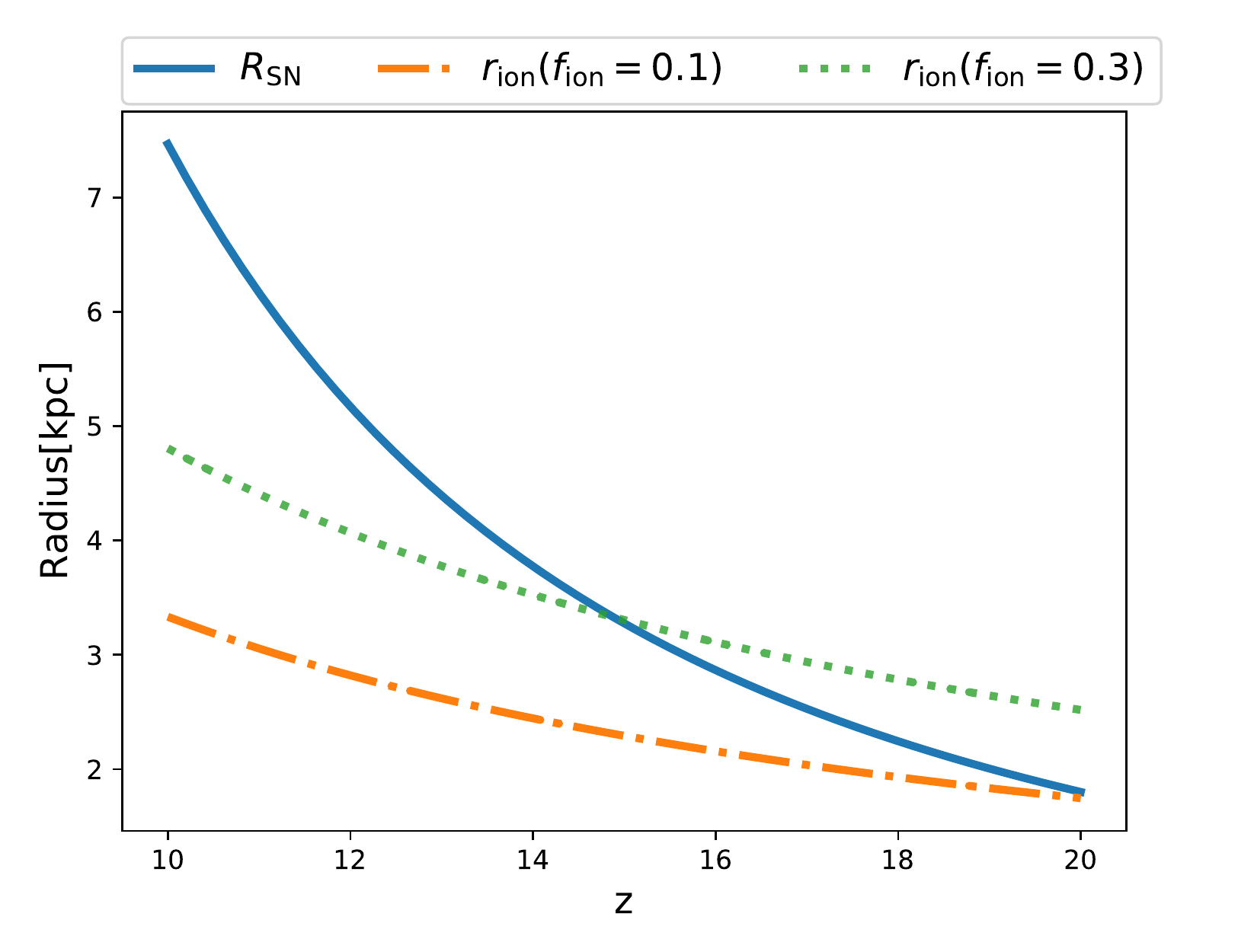}
\caption{The comparison between the Str\"omgren radius in Eq.~\eqref{eq: rion_with_z} and the Sedov-Taylor self-similar solution in Eq.~\eqref{rsn}.}
\label{fig: rion_vs_rsn2}
\end{figure}

Figure~\ref{fig: rion_vs_rsn2} shows the comparison between 
$R_{\rm SN}$ and $r_{\rm ion}$ with two different $f_{\rm esc}$.
The blue solid line shows the redshift evolution of $R_{\rm{SN}}$ in Eq.~\eqref{rsn}, and the orange dotted-dashed and green dotted lines represent the one of radius of the ionized bubble with $f_{\rm{esc}}=0.1$ and $0.3$ respectively.
When $f_{\rm esc} < 0.3$, the figure tells us that 
$R_{\rm{SN}}$ is larger than $r_{\rm ion}$, in
particular, in redshifts (z<15). 
Therefore we can conclude that, in $f_{\rm esc} < 0.3$,
the bubble created by a Pop III star before the SNe
can be destroyed by an SNR and 
SNRs can increase the ionization fraction substantially.
Note that in the above estimation, we assume that the SNR energy does not significantly lose in a dark matter halo. However, 
as the ionizing photons are absorbed in dark matter halos 
and are reduced by $f_{\rm esc}$, some fraction of the SNR energy is consumed
inside a dark matter halo and, then, the SNR radius might be smaller than in Eq.~\eqref{rsn}.
The dependency on the gas density, $r_{\rm s} \propto n_{\rm g}^{-1/3}$ and $R_{\rm SN} \propto n_{\rm g}^{-1/5}$, suggest us that, even in high density $n_{\rm g} \sim 200 n_{\rm g, IGM}$,
the SNR can escape a dark matte halo more easily than the Str\"{o}mgren radius.
Nevertheless considering the propagation of the SNR in a dark matter halo 
requires smaller $f_{\rm esc}$ to satisfy the condition $r_{\rm s} < R_{\rm SN}$.

The overlapping of SNRs also leads to overestimate the SNR contribution to the ionization fraction.
One can see in Fig.~\ref{fig: xe_zeta} that the additional ionization fraction in the early reionization stage due to PISNe is $x_{\rm{e}}^{\rm{SN}}\lesssim \mathcal{O}(0.1)$.
In such a small ionization fraction, the probability of the overlapping would be small. 
However, in massive halos,
there is a possibility that 
the star formation is very effective and many stars form 
almost at the same time.
When such starburst mode happens,
PISNe also happens simultaneously in a massive halo and, resultantly, one large SNR is created with the total energy of all PISNe in this halo.
If this starburst mode is dominant in the star formation,
our constraint would be overestimated.
Although we neglect it, the contribution of small-mass Pop III stars 
also causes the overestimation of the SNR contribution.
The clustering of small-mass Pop III stars near a massive Pop III progenitor of PISNe could create a large bubble before the PISNe.
Therefore, the SNR cannot contribute to increase the ionization fraction when the bubble sized is much larger than the PISNe.
In order to evaluate this effect,
it is required to include the ionized bubble evolution by assuming the IMF, the escape fraction, and clustering of Pop III stars consistently. 
Such computation can be performed in cosmological numerical simulation and it is beyond our scope.

\section{Conclusion}

It is theoretically predicted that massive Pop III stars can cause energetic PISNe at the final stage of their lives.
The generated SNRs expand to several kpc and their inside continues to be fully ionized.
In this paper, we investigate the impact of PISNe of Pop III stars on the reionization history.
The abundance of PISNe is unknown both theoretically and observationally.
Therefore, to model the PISN contribution to cosmic reionization, we have introduced a parameter~$\zeta$, which relates to the abundance of the PISNe to the collapse fraction of the universe.
We have shown that, 
although PISNe cannot ionize the universe entirely enough,
PISNe induce early reionization and its efficiency highly depends on the abundance of PISNe.

Since the early reionization can affect the CMB anisotropies, the CMB anisotropy measurement allows us to obtain the constraint on the abundance of PISNe.
In order to investigate the constraint,
we have performed the MCMC analysis with the latest Planck data incorporating our model of the PISN early reionization.
On top of the PISN contribution,
our reionization model include the conventional "tanh" 
type, which represents the contribution of first galaxies 
and Pop II stars as the main sources of ionization photons.
We have found that when $\zeta < 10^{-3}$, the PISN contribution is totally subdominant, and the constraint on  
the "tanh" type is similar to the constraint without the PISNe.
However, when $\zeta > 10^{-3}$, PISNe strongly affect the Thomson optical depth of CMB and the reionization by "tanh" type delayed to compensate the early reionization due to PISNe.
Our constraint on the PISN abundance is $\zeta <10^{-2}$ from the latest Planck measurement.

In general, the abundance of PISNe depends on the nature of Pop III stars including 
their mass fraction to the dark matter halo mass and
the IMF. 
We have shown that our parameter $\zeta$ is related to
the mass fraction of Pop III stars to dark matter halos of the universe, $f_{\rm star}$, and
the number fraction of PISN progenitors in the total Pop III stars, $f_{\rm mf}$.
Our constraint on $\zeta$ can be converted to $f_{\m{mf}}f_{\rm star} \lesssim 1.4\times10^{-3}$.
Cosmological simulations suggests $f_{\rm star} \sim 10 ^{-3}$ for the Pop III star formation~\cite{2014MNRAS.442.2560W}.
It is difficult to obtain the constraint on the Pop III star IMF,~$f_{\m{mf}}$, from our current analysis.
However, we have also shown that our constraint can be improved and provide useful information on the Pop III star IMF. 
The high redshift QSO observation suggests $x_e \sim 0.5$ at $z \sim 7.5$. When we take into account this result, our constraint can be improved to $\zeta < 10^{-3}$ and $f_{\m{mf}} \lesssim 0.1$. 
Therefore, the further measurements of the ionized fraction in high redshifts allow us to rule out 
the top-heavy IMF, 
in which massive Pop III stars causing PISNe dominate smaller Pop III stars in abundance.

The most effective theoretical uncertainties in our model
is the termination redshift of PISNe, $z_{\m{end}}$. 
Although it strongly depends on the abundance of PISNe, we set $z_{\m{end}}$ by hand. In order to investigate the impact of this redshift on our constraint, we have redone the MCMC analysis for the different redshifts.
We found out that 
the dependence of our constraint on $z_{\m{end}}$ 
is approximated to 
$\m{log}_{10}\zeta \leq -2.0\lr{z_{\m{end}}/12}^{1.22}$
in the range of $10<z_{\m{end}}<15$.

Our constraint is obtained in the isolated SNR assumption with neglecting the reionization due to Pop III stars.
These assumptions are valid in the limited case as discussed in Sec.~\ref{subsec: limit_assumption} and, otherwise,
could lead our result to be overestimated.
Besides, in order to constraint the PISN contribution to the reionization more,
we need to take into account the Pop III star contribution as well.
To address these concerns consistently,
the cosmological numerical simulation could be required.
We leave the detailed study to our future work.

Although our work is based on the optimistic case,
our result illuminates that
the CMB measurement 
has the potential to explore
observational
signatures of PISNe.
Further investigation on the PISNe contribution 
can provide
the access to the nature of Pop III stars.

\acknowledgments
This work is supported by JSPS KAKENHI Grants
No.~JP20J22260 (K.T.A)
and No.~JP21K03533 (H.T)

\bibliography{article}

\end{document}